\documentclass[twocolumn,aps,prc,superscriptaddress,showpacs,nobibnotes]{revtex4}
\usepackage{epsfig,dcolumn,bm}

\newcommand{\lc}{{\cal L}}

\newcommand{\sqd}{\sqrt{2}}

\newcommand{\sqs}{\sqrt{6}}

\begin{document}

\title{Hidden charm dynamically generated resonances and the $e^+e^-\rightarrow J/\psi D \bar D$, $J/\psi D\bar D^*$ reactions}

\author{D. Gamermann}{\thanks{E-mail: daniel.gamermann@ific.uv.es}
\affiliation{Departamento de F\'isica Te\'orica and IFIC, Centro Mixto
Universidad de Valencia-CSIC,\\ Institutos de Investigaci\'on de
Paterna, Aptdo. 22085, 46071, Valencia, Spain} 
\author{E. Oset}{\thanks{E-mail: oset@ific.uv.es}
\affiliation{Departamento de F\'isica Te\'orica and IFIC, Centro Mixto
Universidad de Valencia-CSIC,\\ Institutos de Investigaci\'on de
Paterna, Aptdo. 22085, 46071, Valencia, Spain}

\begin{abstract}
We analyze two recent reactions of Belle, producing $D\bar D$ and $D\bar D^*$ states that have an enhancement of the invariant $D\bar D$, $D\bar D^*$ mass distribution close to threshold, from the point of view that they might be indicative of the existence of a hidden charm scalar and an axial vector meson states below $D\bar D$ or $D\bar D^*$ thresholds, respectively. We conclude that the data is compatible with the existing prediction of a hidden charm scalar meson with mass around 3700 MeV, though other possibilities cannot be discarded. The peak seen in the $D\bar D^*$ spectrum above threshold is, however, unlikely to be due to a threshold enhancement produced by the presence, below threshold, of the hidden charm axial vector meson $X(3872)$.
\end{abstract}



\maketitle

\section{Introduction}

The observation of new states with open and hidden charm has sparked the interest of many experimental and theoretical research groups. In order to accommodate the newly found states, many theoretical explanations have been suggested and there is still controversy about the structure of many of these resonances. New and precise data will help understand these states better and it is important to test the theoretical models against these data. In that sense, the recent reactions of Belle producing $D\bar D$ and $D\bar D^*$ final particles \cite{belle} bring relevant information concerning hidden charm states that we want to exploit in the present work.

The reactions $e^+e^-\rightarrow J/\psi D^{(*)} \bar D^{(*)}$, with different charmed meson pairs, have been recently observed by the Belle collaboration \cite{belle}. We study here the cases with $D\bar D$ and $D\bar D^* +c.c.$ pairs in the final state. The invariant mass distributions for $D$ pair production in these two cases present an important enhancement close to the two meson threshold, which have led to the claim of two new resonance states \cite{belle}.

In three recent papers \cite{hanh,meuax,nml} it was shown that bound states lying close to threshold for meson production can cause a strong enhancement in the cross section above threshold. Moreover, following similar steps as in \cite{lutz1,lutz2,guo1,guo2}, we have developed a model for generating dynamically resonances with open and hidden charm quantum numbers in the scalar and axial sectors \cite{meuax,meusca}. The model describes a large number of observed resonances as dynamically generated states in agreement with \cite{lutz1,lutz2,guo1,guo2}. However, some other states predicted with narrow widths in \cite{lutz1,lutz2,guo1,guo2} appear in \cite{meuax,meusca} as very broad resonances, which could explain why such states are not being observed. In particular the extension of \cite{meuax,meusca} to the hidden charm sector predicts new meson states both scalar and axial vector.

The predicted hidden charm resonances are one scalar and two axial vectors. One of these axials can accommodate the $X(3872)$ \cite{x1,x2} which most probably has positive C-parity conjugation \cite{cpx}, but the other axial pole has negative C-parity and is nearly degenerated in mass with the positive one. The scalar resonance is predicted around 3.7 GeV with a very small width and there is so far no experimental evidence for it. Since the scalar pole lies close to the $D\bar D$ production threshold and the axial ones close to the $D\bar D^*+c.c.$ threshold, it is possible that the processes measured by Belle are dominated by these dynamically generated resonances, so that the threshold enhancement is a consequence of these poles instead of being due to new resonances above threshold.

We briefly present, in the next section, our model for generating dynamically resonances and show how to apply it for calculating cross sections for $J/\psi$ plus $D$ pair production from electron positron annihilation. Section III shows our results and in section IV we present our conclusions.

\section{Framework}

The reaction $e^+e^-\rightarrow J/\psi D\bar D$ can be described by the diagram in fig. \ref{fig1} if one assumes that the $D\bar D$ pair comes from a resonance.

\begin{figure}[h]
\includegraphics[width=6cm]{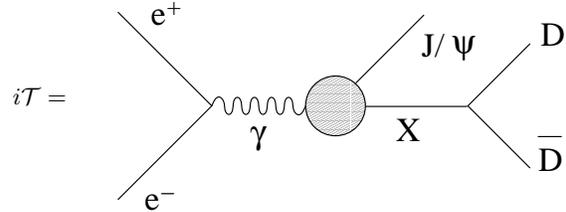} 
\put(-210,40){$i{\cal T}=$}
\caption{Feynman diagram for the process $e^+e^-\rightarrow J/\psi D\bar D$} \label{fig1}
\end{figure}

Close to threshold the only part of this amplitude which is strongly energy dependent is the $X$ propagator and all other parts can be factorized, so that we can write

\begin{eqnarray}
{\cal T}&=&C {1\over M_{inv}^2(D\bar D)-M_X^2+i\Gamma_X M_X} \label{eq1}
\end{eqnarray}
if we describe the $X$ resonance as a Breit-Wigner type.

The cross section would then be given by an integral over the phase space of the three particles in the final state:

\begin{eqnarray}
\sigma&=&{1\over V_{rel}(e^+e^-)} {m_{e^-}\over E_{e^-}} {m_{e^+}\over E_{e^+}} \int{d^3p\over(2\pi)^3}{1\over2E_{J/\psi}(p)} \nonumber \\
&\times& \int{d^3k\over(2\pi)^3}{1\over2E_D(k)} \int{d^3k'\over(2\pi)^3}{1\over2E_{\bar D}(k')} \nonumber\\
&\times& (2\pi)^4\delta(p_{e^+}+p_{e^-}-p-k-k')|{\cal T}|^2 \label{eq2}
\end{eqnarray}

Assuming that $\cal T$ depends only on the $D\bar D$ invariant mass, one can evaluate from eq. (\ref{eq2}) the differential cross section:

\begin{eqnarray}
{d\sigma\over dM_{inv}(D\bar D)}&=&{1\over(2\pi)^3}{m_e^2\over s\sqrt{s}} |\overrightarrow k| |\overrightarrow p| |{\cal T}|^2 \label{dcross}
\end{eqnarray}
where $s$ is the center of mass energy of the electron positron pair squared and $|\overrightarrow k|$ and $|\overrightarrow p|$ are given by:

\begin{eqnarray}
|\overrightarrow p|&=&{\lambda^{1/2}(s,M_{J/\psi}^2,M_{inv}^2(D\bar D))\over2\sqrt{s}} \label{pmom} \\
|\overrightarrow k|&=&{\lambda^{1/2}(M_{inv}^2(D\bar D),M_{D}^2,M_{\bar D}^2)\over2M_{inv}(D\bar D)} \label{kmom}
\end{eqnarray}
 Where $\lambda^{1/2}(s,m^2,M^2)$ is the usual K\"allen function.

Let us briefly see how the approach of \cite{meuax,meusca} generates dynamically some resonances. The model starts by writing a Lagrangian for the interaction of the meson fields belonging to the 15-plet of $SU(4)$. For the pseudoscalar field we take 

\begin{widetext}
\begin{eqnarray}
\Phi&=&\sum_{i=1}^{15}{\varphi_i \over \sqd}\lambda_i \nonumber = \left( \begin{array}{cccc}
{\pi^0 \over \sqd}+{\eta \over \sqs}+{\eta_c \over \sqrt{12}} & \pi^+ & K^+ & \bar D^0 \\ & & & \\
\pi^- & {-\pi^0 \over \sqd}+{\eta \over \sqs}+{\eta_c \over \sqrt{12}} & K^0 & D^- \\& & & \\
K^- & \bar K^0 & {-2\eta \over \sqs}+{\eta_c \over \sqrt{12}} & D_s^- \\& & & \\
D^0 & D^+ & D_s^+ & {-3\eta_c \over \sqrt{12}} \\ \end{array} \right) \label{phig},
\end{eqnarray}
\end{widetext}
and an analog one for the vector meson field.

For each one of these fields a current ($J_\mu$ and $\cal J_\mu$) is constructed and the Lagrangians are build by connecting these two currents:

\begin{eqnarray}
\lc_{PPPP}&=&{1\over12f^2}Tr({J}_\mu {J}^\mu+\Phi^4 M) \label{lag} \\
\lc_{PPVV}&=&{-1\over 4f^2}Tr\left(J_\mu\cal{J}^\mu\right). \label{lagini}
\end{eqnarray}

The Lagrangians in (\ref{lag}) and (\ref{lagini}) are $SU(4)$ invariant, but since this is not an exact symmetry in Nature, one breaks it by suppressing the terms in these Lagrangians where the underlying interaction is driven by the exchange of a heavy vector meson. A different $SU(4)$ symmetry breaking pattern borrowed from \cite{skyrme} was also studied in \cite{meusca} in order to estimate the uncertainties of the model.

The tree level amplitudes for meson meson scattering are obtained from the Lagrangians for all possible processes. All meson pairs forming a system with equal quantum numbers span a coupled channel space. All amplitudes, projected in s-wave, for the processes in each coupled channel space are collected into a matrix. We call this matrix the potential, $V$.

The potential is then used to solve the Bethe-Salpeter equation that, in an on-shell formalism \cite{noverd,oller}, assumes an algebraic form:

\begin{eqnarray}
T&=&(\hat 1-VG)^{-1}V \label{bseq1} \\
T&=&-({\hat 1} + V{\hat G})^{-1}V \overrightarrow{\epsilon}.\overrightarrow{\epsilon}' \label{bseq2}
\end{eqnarray}
Equation (\ref{bseq1}) is used for calculating the T-matrix in the scattering of two pseudoscalars that, in s-wave have the quantum numbers of a scalar resonance. Equation (\ref{bseq2}) is used in the scattering of pseudoscalars with vector mesons, takes into account the polarization of the vector mesons and the resulting states have quantum numbers of an axial. The matrix $G$ is diagonal, coming from the loop function of the two meson propagators of the intermediate states.

The loop function has its imaginary part fixed in order to ensure unitarity of the T-matrix. Resonances are identified as poles in the second Riemann sheet of the T-matrix.

With this model many known states have been reproduced: a scalar antitriplet with charm quantum number that has been identified with the $D_{s0}^*(2317)$ and $D_0^*(2400)$ states and two axial antitriplets identified with the $D_{s1}(2460)$, $D_1(2430)$, $D_{s1}(2536)$ and $D_1(2420)$. Apart from those, some broad sextets are predicted in the scalar and axial sectors, though the ones in the scalar sector are probably too broad to have any experimental relevance. The light scalars and the light axials are also reproduced in our model, and finally in the hidden charm sector three poles are generated in the T-matrix, one scalar around 3.7 GeV and two axials, nearly degenerated in mass, with opposite parity and close to the $D\bar D^*$ threshold, one of which is identified with the $X(3872)$.

Since the dynamically generated states are characterized by poles appearing in the unitary T-matrices, the dynamics of our approach is incorporated in the $e^+e^-\rightarrow J/\psi D\bar D$ process by substituting the Breit-Wigner amplitude of (\ref{eq1}) by the $D\bar D$ T-matrix calculated from eq. (\ref{bseq1}). For the reaction $e^+e^-\rightarrow J/\psi D\bar D^*$ everything is done analogously using eq. (\ref{bseq2}) for the T-matrix.

In the following section we explain how we compare our results to Belle's data.

\section{Results}

Belle has measured the differential cross section for $J/\psi D\bar D$, $J/\psi D\bar D^*$ and $J/\psi D^*\bar D^*$ production from electron positron collision at center of mass energy $\sqrt{s}$=10.6 GeV \cite{belle}. We are going to study the first two cases, where the hidden charm states generated in the model of \cite{meuax,meusca} could be related to. The Belle's measurement produces invariant mass distributions for the $D\bar D$ and $D\bar D^*$ that range from threshold up to 5.0 GeV. Our model is, in principle reliable for energies within few hundreds of MeV from the thresholds, so we are going to compare numerically our results with the data up to 4.2 GeV.

The experiment measures counts per bin. In the case of a $D\bar D$ pair, the bins have 50 MeV width, while for the $D\bar D^*$ pair they have 25 MeV. To compare the shape of our theoretical calculation with the experimental data we integrate our theoretical curve in bins of the same size as the experiment and normalize our results so that the total integral of our curve matches the total number of events measured in the invariant mass range up to 4.2 GeV.

The comparison is made by the standard $\chi^2$ test. The value of $\chi^2$ divided by the number of degrees of freedom is given by:

\begin{eqnarray}
{\chi^2 \over d.o.f.}&=&{1\over N}\sum_1^N {(Y_{theo}-Y_{exp})^2\over (\Delta Y_{exp})^2}
\end{eqnarray}
where $Y$ is the number of counts in each bin, $\Delta Y$ is the experimental uncertainty in each measurement and $N$ is the total number of points.

As described in \cite{meusca,meuax}, in the heavy sector the model has one free parameter, $\alpha_H$ which is the subtraction constant in the loop for channels with at least one heavy particle. In these previous papers this parameter has been fitted so that the pole in the C=1, S=1, I=0 sector matches the observed state with these quantum numbers ($D_{s0}(2317)$ and $D_{s1}(2460)$ for scalar and axial states, respectively). The channels in this sector have always one heavy and one light meson, and in principle one could fit a different $\alpha$ for channels with two heavy particles. Here we are going to present results for different values of $\alpha$ in channels with hidden charm (doubly heavy channels). Since we are working with the C=0 sector, we have also channels involving only light mesons. These have negligible influence in the pole position of the hidden charm poles as shown in \cite{meusca}, so we leave $\alpha_L$ constant.

In table \ref{tab1} we show results, for different values of $\alpha_H$, of the pole position of the hidden charm resonance in the scalar sector, and the value of $\chi^2$ calculated with the data from Belle, with combinatorial background already subtracted, for all points below 4.2 GeV in the $J/\psi D\bar D$ production. Fig. \ref{fig2} shows plots of our theoretical histograms compared with experimental data. Note that although we are plotting all points until 5.0 GeV, only the ones below 4.2 have been used in the calculation of $\chi^2$ and in the normalization of the theoretical curves.

\begin{table}
\begin{center}
\caption{Results of $M_X$ and $\chi^2$ for different values of $\alpha_H$.} \label{tab1}
\begin{tabular}{c|cc}
\hline
$\alpha_H$ & $M_X$(MeV) & $\chi^2 \over d.o.f$ \\
\hline
\hline
-1.4   & 3701.93-i0.08 & 0.38 \\
-1.3   & 3718.93-i0.06 & 0.36 \\
-1.2  & 3728.12-i0.03 & 0.49 \\
-1.1   & Cusp & 0.66 \\
\hline
\end{tabular}
\end{center}
\end{table}

\begin{figure}[h]
\begin{center}
\includegraphics[width=6.5cm,angle=-90]{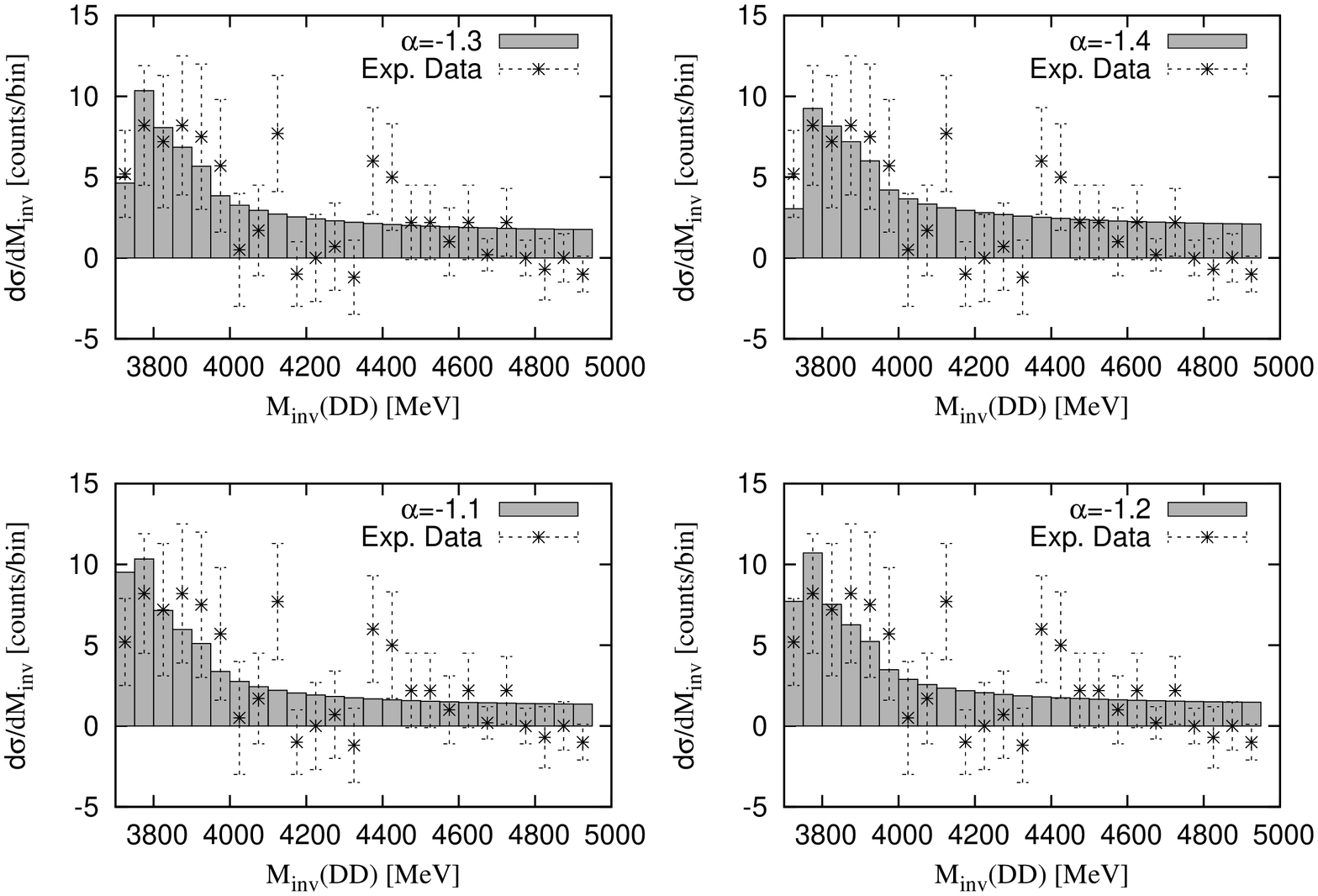} 
\caption{Theoretical histograms compared with data from \cite{belle} for $D\bar D$ invariant mass distribution.} \label{fig2}
\end{center}
\end{figure}

The $\chi^2$ values obtained in table \ref{tab1} are all below 1, indicating a good fit to the data in all curves. This is in part due to the large experimental errors, but the clear message is that the presence of a pole below $D\bar D$ threshold is enough to produce the observed enhancement of the cross section for this reaction in the $D\bar D$ invariant mass above threshold. The results of table \ref{tab1} and inspection of fig. \ref{fig2} show some preference for values of $\alpha_H$=-1.3, -1.4, which would correspond to the hidden charm scalar with mass slightly above 3700 MeV.

For the production of $J/\psi D\bar D^*$ we use the model for generating axial resonances. In this case the resonance $X$ in fig. \ref{fig1} should be identified with the $X(3872)$ generated by our model. Note that our predicted state with negative C-parity does not fit here, since this experiment selects a positive C-parity state for the $X$. Table \ref{tab2} shows results for $M_X$ and $\chi^2$ for different values of $\alpha_H$. Since the state $X(3872)$ is known and has a rather precise mass, we have chosen a smaller range to vary the parameter $\alpha$ in order to get the mass of the $X$ closer to its experimental value. Fig. \ref{fig3} compares our theoretical results with the experimental data from Belle.

\begin{table}
\begin{center}
\caption{Results of $M_X$ and $\chi^2$ for different values of $\alpha_H$.} \label{tab2}
\begin{tabular}{c|cc}
\hline
$\alpha_H$ & $M_X$(MeV) & $\chi^2 \over d.o.f$ \\
\hline
\hline
-1.40   &3865.09-i0.00& 2.84 \\
-1.35   &3870.07-i0.00& 3.84 \\
-1.30  & 3872.67-i0.00 & 5.04 \\
-1.25   & Cusp & 5.01 \\
\hline
\end{tabular}
\end{center}
\end{table}

\begin{figure}[h]
\begin{center}
\includegraphics[width=6.5cm,angle=-90]{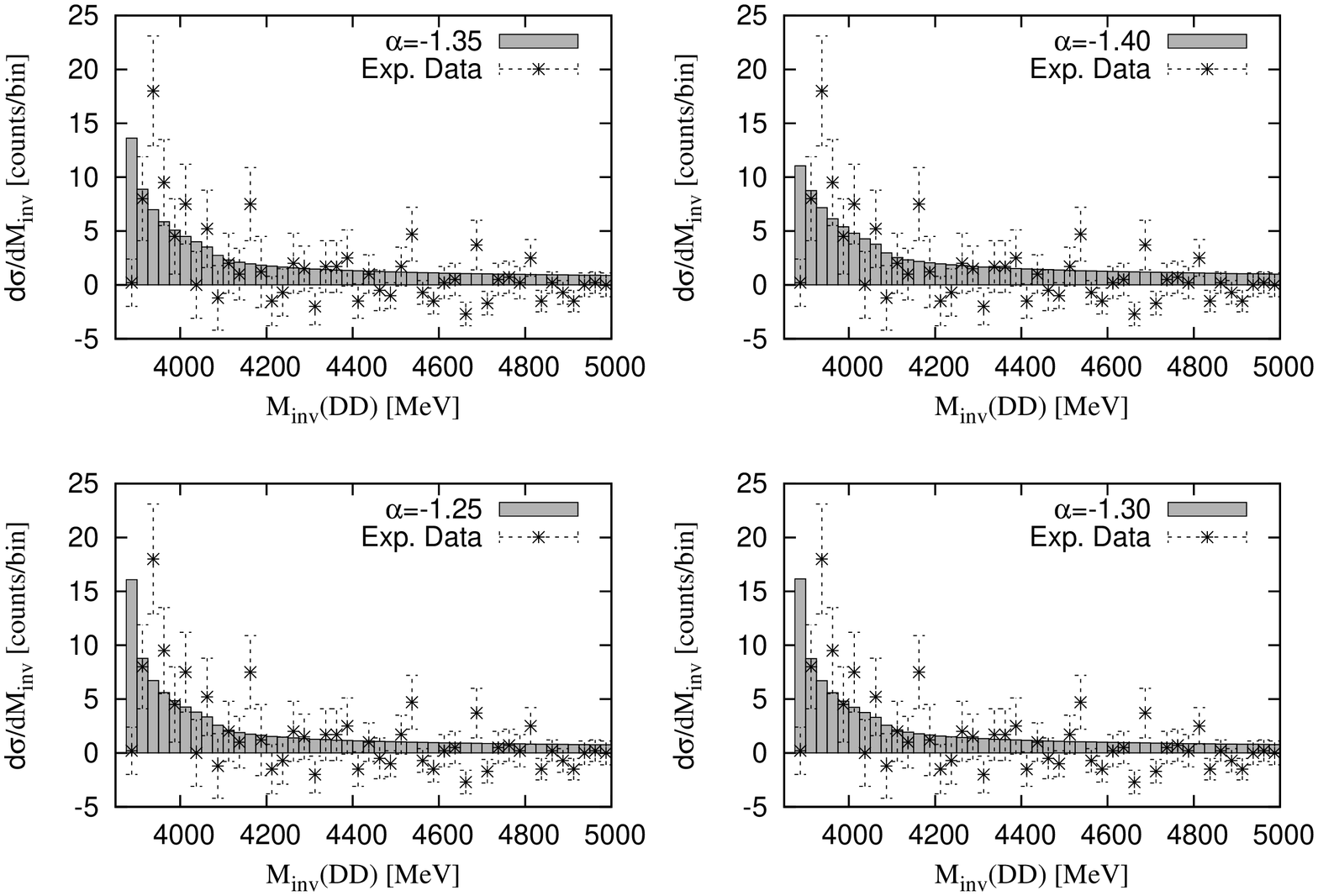} 
\caption{Theoretical histograms compared with data from \cite{belle} for $D\bar D^*$ invariant mass distribution.} \label{fig3}
\end{center}
\end{figure}

In this case the $\chi^2$ obtained is in all cases of the order of 3 or higher, clearly indicating a poor fit to the data.

The peaks seen in the experiment have been fitted with Breit-Wigner like resonances in \cite{belle}. In the experimental work of \cite{belle} the two peaks discussed here have been fitted in terms of Breit-Wigner amplitudes, suggesting two new resonances. In order to make the results obtained here more meaningful, we also perform such a fit and compare the results. We take the same Breit-Wigner parameters suggested in the experimental paper. The scalar resonance has $M_X$=3878 MeV and $\Gamma_X$=347 MeV and the axial one has $M_X$= 3942 MeV and $\Gamma_X$= 37 MeV. We show the results obtained by fitting a Breit-Wigner form from eq. (\ref{eq1}) in $\cal T$ of eq. (\ref{dcross}) in figs. \ref{fig4}, \ref{fig5}. Additionally we calculate $\chi^2$ and find $\chi^2/d.o.f$=1.04 for the $D\bar D$ distribution and $\chi^2/d.o.f$=1.03 for the $D\bar D^*$ distribution. The value of $\chi^2$ for the $D\bar D$ distribution can be improved if we take different parameters for the Breit-Wigner resonance. Taking for the fit $M_X$=3750 MeV and $\Gamma_X$=250 MeV we obtain a value of $\chi^2/d.o.f$=0.55, comparable with those obtained in our analysis. The value of $\chi^2$ for the $D\bar D^*$ distribution is undoubtedly better in the case of a Breit-Wigner fit that in our analysis assuming the $X(3872)$ resonance as responsible for the peak.

As a consequence of the discussion, our conclusions would be a support for a new resonance around 3940 MeV as suggested in \cite{belle}, while for the case of the broad peak seen in $D\bar D$, the weak case in favor of a new state around 3880 MeV discussed in \cite{belle} is further weakened by the analysis done here, showing that the results are compatible with the presence of a scalar hidden charm state with mass around 3700 MeV.

\begin{figure}[h]
\includegraphics[width=6cm,angle=-90]{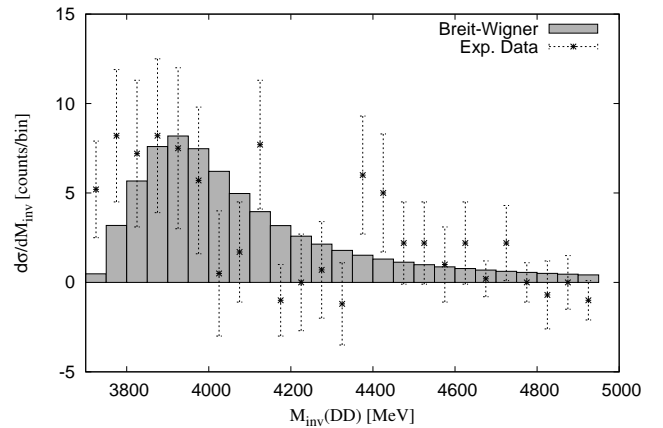} 
\caption{Histograms calculated with Breit-Wigner resonance with mass $M_X$=3880 MeV compared to data.} \label{fig4}
\end{figure}

\begin{figure}[h]
\includegraphics[width=6cm,angle=-90]{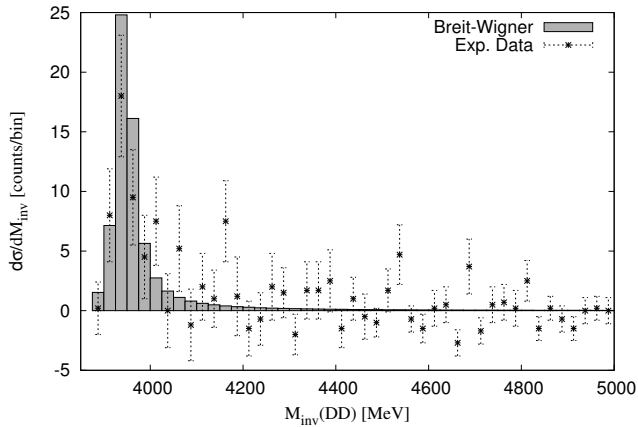} 
\caption{Histograms calculated with Breit-Wigner resonance with mass $M_X$=3940 MeV compared to data.} \label{fig5}
\end{figure}

\section{Conclusions}

We have applied our model for generating dynamically resonances with hidden charm in order to describe differential cross sections in reactions observed by Belle where $D$ meson pairs were produced together with $J/\psi$ from electron positron collisions at a center of mass energy of 10.6 GeV.

In the experiment two peaks are observed, one in the $D\bar D$ invariant mass distribution around 3880 MeV and another in the $D \bar D^*$ invariant mass distribution around 3940 MeV. We studied here the possibility that these peaks might be a threshold enhancement caused by resonances below the threshold for $D$ pair production. 

While our theoretical model describes very well the peak in the $D\bar D$ invariant mass as caused by a scalar resonance which is mainly a $D\bar D$ bound state, the $\chi^2$ values obtained in describing the peak in the $D\bar D^*$ invariant mass distribution are too big, indicating that this peak is unlikely to be caused by the $X(3872)$ resonance reproduced in our approach as a dynamically generated resonance.

The results obtained for the invariant mass distribution calculated with our model in comparison with data are compatible with the existence of the predicted hidden charm scalar resonance below $D\bar D$ threshold, but the large experimental errors do not discard other possibilities. Further experiments to search for this predicted states should be most welcome. In particular the radiative decay of this predicted state leading to $J/\psi \gamma$ studied in \cite{meurad} would be a good test for it.

\acknowledgements{We would like to thank Dr. P. Pakhlov for making Belle's data accessible for us and for answering our questions about the measurements and the experiment. One of us, D. Gamermann, wishes to acknowledge support from the Ministerio de Educacion y Ciencia in the FPI program. This work was supported in part by DGICYT contract number FIS2006-03438 and the Generalitat Valenciana. This research is  part of the EU Integrated Infrastructure Initiative Hadron Physics Project under contract number RII3-CT-2004-506078.


\begin{thebibliography}{99}

\bibitem{belle}
   K.~Abe {\it et al.}  [Belle Collaboration],
  arXiv:0708.3812 [hep-ex].

\bibitem{hanh}
  C.~Hanhart, Yu.~S.~Kalashnikova, A.~E.~Kudryavtsev and A.~V.~Nefediev,
  arXiv:0704.0605 [hep-ph].


\bibitem{meuax}
 D.~Gamermann and E.~Oset,
  Eur.\ Phys.\ J.\  A {\bf 33}, 119 (2007)
  [arXiv:0704.2314 [hep-ph]].


\bibitem{nml}
  E.~Braaten and M.~Lu,
  arXiv:0709.2697 [hep-ph].

\bibitem{meusca}
  D.~Gamermann, E.~Oset, D.~Strottman and M.~J.~Vicente Vacas,
  Phys.\ Rev.\  D {\bf 76}, 074016 (2007)
  [arXiv:hep-ph/0612179].

\bibitem{lutz1}
 E.~E.~Kolomeitsev and M.~F.~M.~Lutz,
  Phys.\ Lett.\ B {\bf 582} (2004) 39
  [arXiv:hep-ph/0307133].

\bibitem{lutz2}
 J.~Hofmann and M.~F.~M.~Lutz,
  Nucl.\ Phys.\ A {\bf 733}, 142 (2004)
  [arXiv:hep-ph/0308263].

\bibitem{guo1}
 F.~K.~Guo, P.~N.~Shen and H.~C.~Chiang,
  Phys.\ Lett.\  B {\bf 647}, 133 (2007)
  [arXiv:hep-ph/0610008].

\bibitem{guo2}
 F.~K.~Guo, P.~N.~Shen, H.~C.~Chiang and R.~G.~Ping,
  Phys.\ Lett.\ B {\bf 641} (2006) 278
  [arXiv:hep-ph/0603072].

\bibitem{x1}
  S.~K.~Choi {\it et al.}  [Belle Collaboration],
  Phys.\ Rev.\ Lett.\  {\bf 91}, 262001 (2003)
  [arXiv:hep-ex/0309032].

\bibitem{x2}
  D.~Acosta {\it et al.}  [CDF II Collaboration],
  Phys.\ Rev.\ Lett.\  {\bf 93}, 072001 (2004)
  [arXiv:hep-ex/0312021].

\bibitem{cpx}
 K.~Abe {\it et al.},
  arXiv:hep-ex/0505037.

\bibitem{skyrme}
  H.~Walliser,
  Nucl.\ Phys.\  A {\bf 548}, 649 (1992).

\bibitem{noverd}
  J.~A.~Oller and E.~Oset,
  Phys.\ Rev.\  D {\bf 60}, 074023 (1999)
  [arXiv:hep-ph/9809337].

\bibitem{oller}
  J.~A.~Oller and U.~G.~Meissner,
  Phys.\ Lett.\  B {\bf 500}, 263 (2001)
  [arXiv:hep-ph/0011146].

\bibitem{meurad}
 D.~Gamermann, L.~R.~Dai and E.~Oset,
  arXiv:0709.2339 [hep-ph].

\end{thebibliography}
\end{document}